\documentclass[a4paper,11pt]{article}
\usepackage{pos}
\usepackage{gensymb}
\usepackage{multirow}
\usepackage{caption}
\usepackage{subcaption}

\newcommand{\Xmax}{$X_\textrm{max}$}

\title{Reconstruction procedure of the Fluorescence detector Array of Single-pixel Telescopes (FAST)}
\ShortTitle{Reconstruction procedure of FAST}

\author*[a]{Fraser~Bradfield}
\author[b]{Justin~Albury}
\author[b]{Jose~Bellido}
\author[c]{Ladislav~Chytka}
\author[d]{John~Farmer}
\author[a]{Toshihiro~Fujii}
\author[e]{Petr~Hamal}
\author[e]{Pavel~Horvath}
\author[e]{Miroslav~Hrabovsky}
\author[e]{Vlastimil~Jilek}
\author[e]{Jakub~Kmec}
\author[e]{Jiri~Kvita}
\author[d]{Max~Malacari}
\author[c]{Dusan~Mandat}
\author[f]{Massimo~Mastrodicasa}
\author[g]{John~N.~Matthews}
\author[e]{Stanislav~Michal}
\author[h]{Hiromu~Nagasawa}
\author[h]{Hiroki~Namba}
\author[e]{Libor~Nozka}
\author[c]{Miroslav~Palatka}
\author[c]{Miroslav~Pech}
\author[d]{Paolo~Privitera}
\author[a]{Shunsuke~Sakurai}
\author[f]{Francesco~Salamida}
\author[c]{Petr~Schovanek}
\author[d]{Radomir~Smida}
\author[e]{Daniel~Stanik}
\author[e]{Zuzana~Svozilikova}
\author[i]{Akimichi~Taketa}
\author[h]{Kenta~Terauchi}
\author[g]{Stan~B.~Thomas}
\author[c,e]{Petr~Travnicek}
\author[c]{Martin~Vacula}

\affiliation[a]{Graduate School of Science, Osaka Metropolitan University, Sumiyoshi-ku, Osaka, Japan}
\affiliation[b]{Department of Physics, University of Adelaide, Adelaide, S.A., Australia}
\affiliation[c]{Institute of Physics of the Academy of Sciences of the Czech Republic, Prague, Czech Republic}
\affiliation[d]{Kavli Institute for Cosmological Physics, University of Chicago, Chicago, IL, USA}
\affiliation[e]{Joint Laboratory of Optics of PU and IF of CAS, Palacky University, Olomouc, Czech Republic}
\affiliation[f]{Department of Physical and Chemical Sciences, University of L’Aquila and INFN LNGS}
\affiliation[g]{High Energy Astrophysics Institute and Department of Physics and Astronomy, University of Utah, Salt Lake City, UT, USA}
\affiliation[h]{Graduate School of Science, Kyoto University, Sakyo-ku, Kyoto, Japan}
\affiliation[i]{Earthquake Research Institute, University of Tokyo, Bunkyo-ku, Tokyo, Japan}

\emailAdd{sw22383p@st.omu.ac.jp}

\abstract{The Fluorescence detector Array of Single-pixel Telescopes (FAST) is one of several proposed designs for a next-generation cosmic-ray detector. Such detectors will require enormous collecting areas whilst also needing to remain cost-efficient. To meet these demands, the FAST collaboration has designed a simplified, low-cost fluorescence telescope consisting of only four photomultiplier tubes (PMTs). Since standard air shower reconstruction techniques cannot be used with so few PMTs, FAST utilises an alternative two-step approach. In the first step, a neural network is used to provide a first estimate of the true shower parameters. This estimate is then used as the initial guess in a minimisation procedure where the measured PMT traces are compared to simulated ones, and the best-fit shower parameters are found. A detailed explanation of these steps is given, with the expected performance of FAST prototypes at the Telescope Array experiment acting as a demonstration of the technique.}

\ConferenceLogo{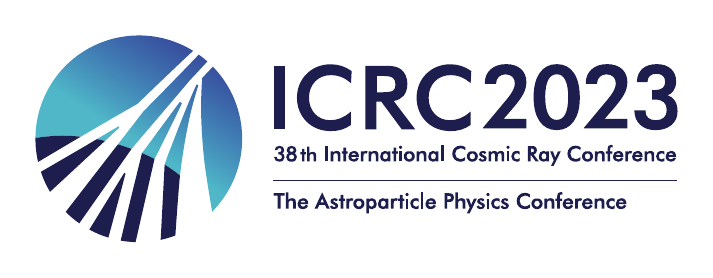}

\FullConference{%
38th International Cosmic Ray Conference (ICRC2023)\\
  26 July - 3 August, 2023\\
  Nagoya, Japan}

\begin{document}
\maketitle

\section{Ultra-high energy cosmic rays and FAST}
The origin of ultra-high energy cosmic rays (UHECRs) remains an open problem, one that is being tackled from several different angles in the current age of multi-messenger astronomy \cite{Coleman:2022abf}. For experiments aiming to observe UHECRs as they arrive at Earth, the steep nature of the cosmic ray energy spectrum necessitates the construction of observatories stretching across hundreds to thousands of square kilometers. 

Currently, the Pierre Auger Observatory (Auger) in Mendoza, Argentina \cite{bib:auger}, and the Telescope Array Experiment (TA) in Utah, USA \cite{bib:tasd, bib:tafd}, are the largest cosmic ray observatories in the world. These experiments measure the showers of secondary particles produced when cosmic rays interact with atmospheric molecules in the upper atmosphere, so-called `extensive air showers' (EASs). Both experiments utilise a hybrid approach to measure EASs, precisely reconstructing the energy, arrival direction and mass composition information of primary cosmic rays. Fluorescence telescopes measure the fluorescence light emitted when EAS particles interact with atmospheric nitrogen/oxygen, whilst ground based particle detectors, such as water Cherenkov stations or plastic scintillation detectors, measure the secondary particle density at ground level.

\begin{figure}[h]
    \centering
    \includegraphics[width=\textwidth]{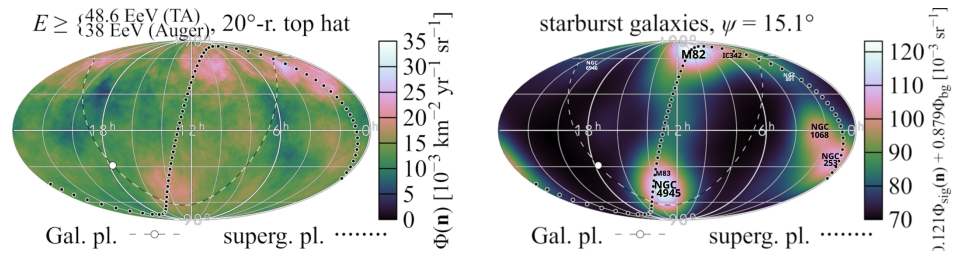}
    \caption{(Left) Observed UHECR flux above 38\,EeV (49\,EeV on the TA scale) by Auger and TA. (Right) UHECR flux predicted by the starburst galaxy model.}
    \label{fig:sbg}
\end{figure}

Recent results from the Auger-TA working group have shown a correlation between arrival directions of UHECRs with $E>38$\,EeV and starburst galaxies \cite{PierreAugerTelescopeArray:2023workinggroup}. Figure \ref{fig:sbg} shows a comparison between the observed flux and the model prediction which uses a catalog of starburst galaxies between 1-130\,Mpc. The post trial significance is 4.2$\sigma$ .

Searches for anisotropy and/or potential sources at even higher energies, where deflections from initial trajectories by intervening magnetic fields are less (for a given charge), are limited by low statistics. To make such studies viable, new observatories will require collecting areas more than an order of magnitude greater than current experiments. The Fluorescence detector Array of Single-pixel Telescopes (FAST) aims to address this requirement by utilising simplified, low-cost, easily deployable fluorescence telescopes installed over a targeted area of 150,000\,km$^2$ \cite{bib:fast, bib:fast2}. 

\begin{figure}
    \centering
    \includegraphics[width=0.75\textwidth]{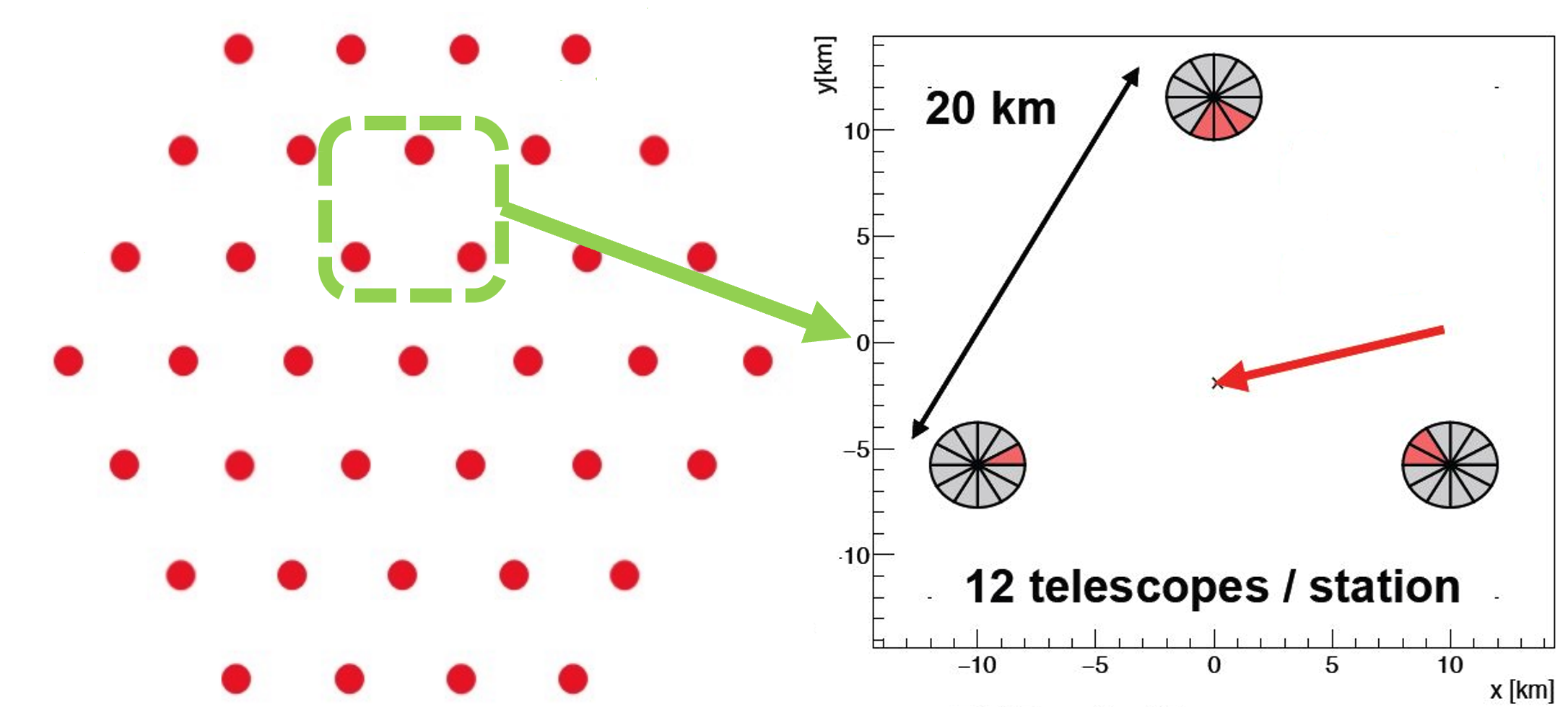}
    \caption{Schematic layout of a scaled down FAST array. The red arrow in the right image shows the arrival direction and core location of a simulated shower. FAST telescopes measuring a significant signal for this event are highlighted in red.}
    \label{fig:FAST}
\end{figure}

The FAST telescope design consists of four 20\,cm photomultiplier-tubes (PMTs) covering a $30\degree \times 30\degree$ field-of-view, located at the focal plane of a 1.6\,m diameter segmented mirror \cite{bib:fastoptics}. A full-sized FAST array will consist of 500 `stations', where each station contains 12 FAST telescopes and covers a full $360\degree$ in azimuth and $30\degree$ in elevation. Stations will be arranged in a triangular pattern with a spacing of 20\,km. Figure \ref{fig:FAST} shows a schematic of the proposed layout.

\section{FAST reconstruction procedure - overview}
A precise and accurate reconstruction of observed events is critical for FAST to achieve its goal of extending cosmic ray measurements beyond 10$^{20}$\,eV. Standard fluorescence telescope reconstruction methods, such as used by Auger and TA, use finely pixelated cameras (several hundreded PMTs) to determine the geometry and energy deposit profile of detected showers \cite{bib:augerFD, bib:tafd}. However, with only four PMTs, FAST telescopes are unable to make use of these techniques. Instead, FAST primarily relies on a `top-down' reconstruction algorithm, which compares simulated traces to measured data and maximises a likelihood function to select the best-fit parameters. The fitted parameters are energy, depth of shower maximum \Xmax, zenith, azimuth and core location (two parameters). Other parameters describing the shape of the Gaisser-Hillas function are fixed to constant values in the fitting procedure, since it is expected that FAST measurements will not be sensitive enough to reconstruct them. Previous work has shown that the top-down reconstruction requires an initial guess `close' to the true shower parameters to obtain accurate results \cite{Thomas-Albury:2020thesis}. For this, we make use of machine learning tools which are able to create a sufficiently accurate mapping between the PMT signals and the six desired outputs. A detailed breakdown of both the machine learning and top-down reconstruction components are given in the following sections.

\section{Machine learning based first-guess estimation}
\label{sec:ml}
The first step in the FAST reconstruction chain is to obtain a first estimate of the shower parameters. Whilst exactly how close this estimate must be to achieve a `successful' reconstruction is yet to be quantified, utilising a simple, feed-forward, fully-connected neural network with a triangular array of FAST stations (as depicted in Figure \ref{fig:FAST}) has given promising results. In \cite{Thomas-Albury:2020thesis}, the neural network approach was developed and shown to give initial estimates able to be optimized by the top-down reconstruction for three representative core positions. Work performed in \cite{FAST:2021toshi} extended the range of core positions, achieving resolutions of 4.2$\degree$ in arrival direction, 465\,m in core position, 8\% in energy, and 30\,gcm$^{-2}$ in \Xmax{} at 40 EeV for 3-fold coincidences. 

\begin{figure}[t]
    \centering
    \includegraphics[width=0.8\textwidth]{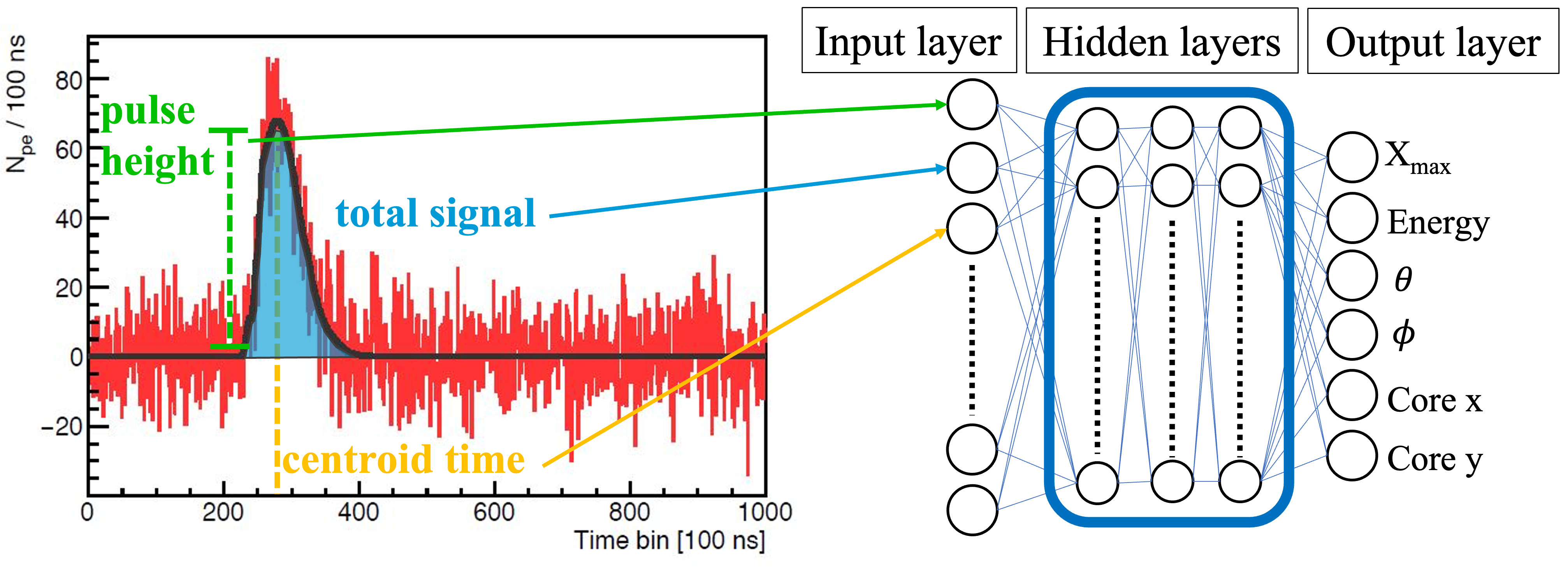}
    \caption{Overview of the machine learning based, first guess estimation. Input features are calculated based on the red trace, which is the simulated black trace with added noise.}
    \label{fig:ml}
\end{figure}

These models were trained on simulated data where the input features are the centroid time, pulse height and integrated signal for each PMT. To increase training efficiency all parameters are normalised before being input into the network. The pulse heights and integrated signals are normalised by dividing by their respective averages over the entire data set. The centroid times are normalised with respect to the arrival time of the earliest signal in a given event and the standard deviation of centroid times over the whole data set. Approximately 500,000 showers are used in training with \Xmax{} and energy values sampled uniformly in the ranges 500--1200\,gcm$^{-2}$ and 1--100\,EeV respectively. Zenith and azimuth values are sampled so as to uniformly populate a hemisphere. The models are constructed using TensorFlow and Keras in Python with the mean squared errors loss function, `Adam' optimizer and 3 hidden layers. ReLU is used as the activation function. The total parameter count is on the order of 10$^6$. Finally, events with no PMT recording a signal with a signal-to-noise ratio (SNR) $>5\sigma$ are excluded from training. Figure \ref{fig:ml} shows the basic structure of these models.

\begin{figure}[h]
     \centering
     \begin{subfigure}[b]{0.34\textwidth}
         \centering
         \includegraphics[width=\textwidth]{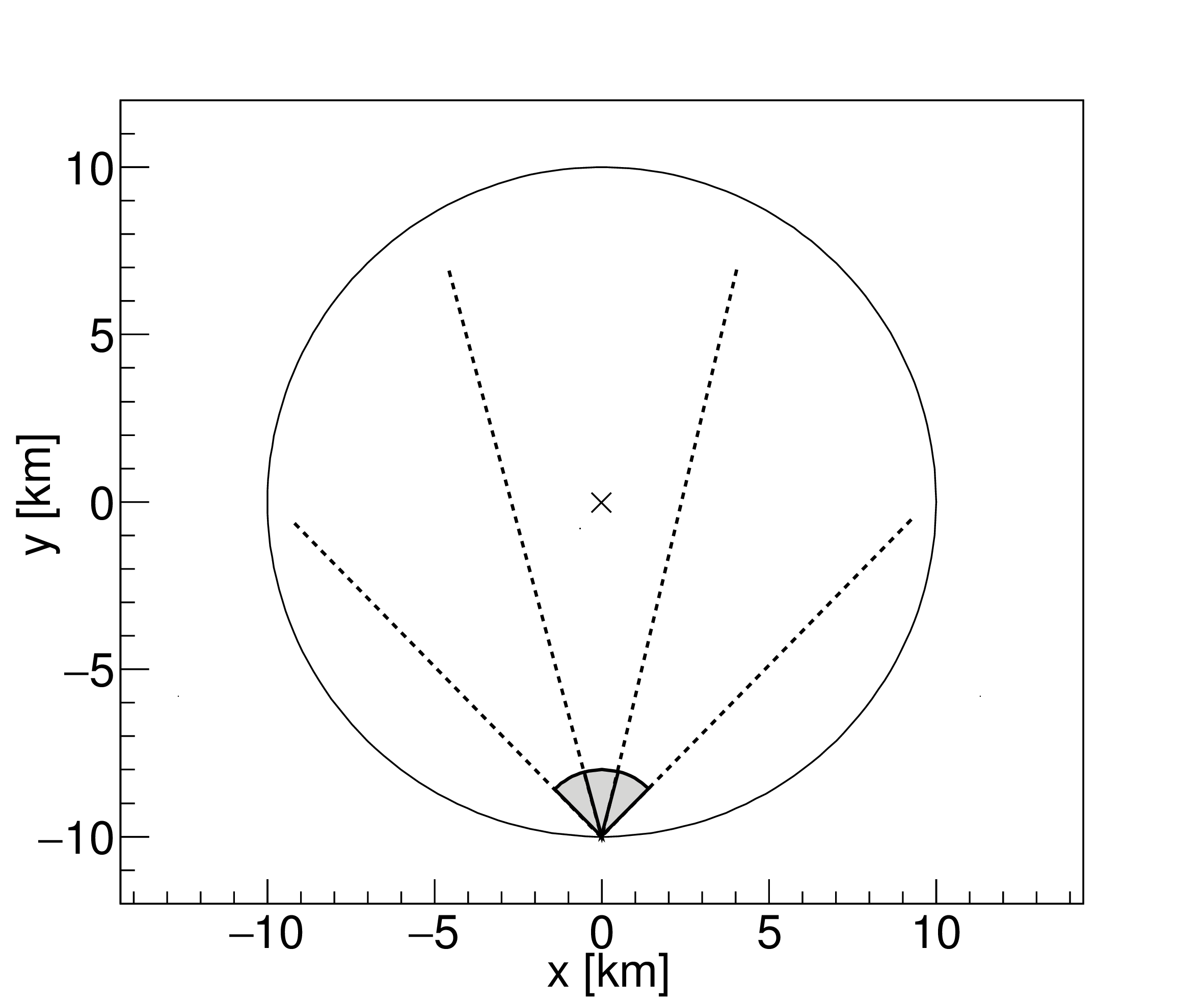}
         \caption{}
         \label{fig:talayout}
     \end{subfigure}
     \begin{subfigure}[b]{0.65\textwidth}
         \centering
         \includegraphics[width=0.49\textwidth]{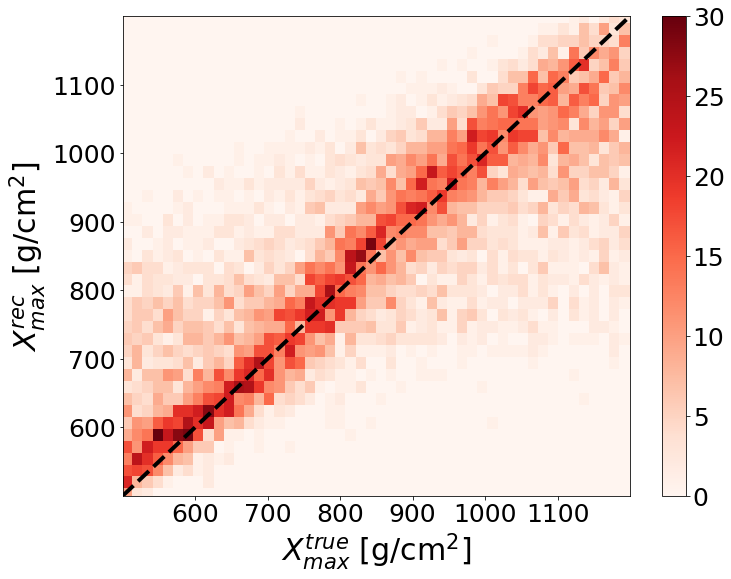}
         \includegraphics[width=0.48\textwidth]{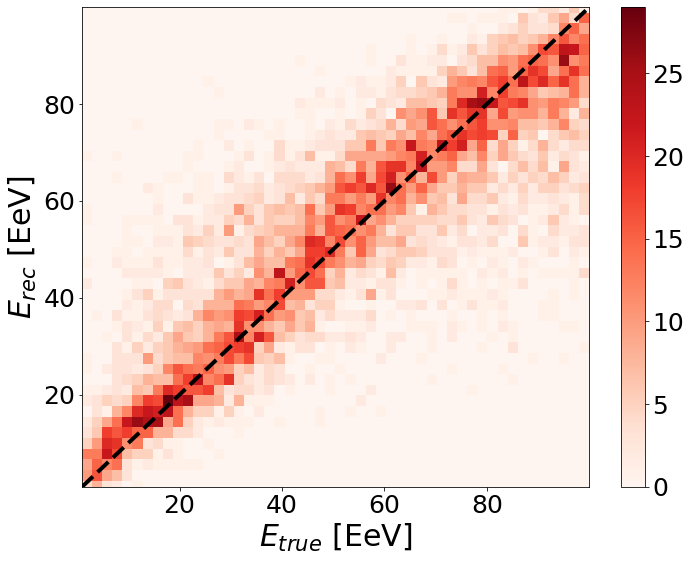}
         \caption{}
         \label{fig:mlresult}
     \end{subfigure}
    \caption{(a) Layout of FAST@TA. The dotted lines show the fields of view of each telescope. The circle indicates the range of core positions used during training. (b) Reconstructed values of \Xmax{} and energy as predicted by the FAST@TA neural network trained with additional geometric inputs.}
    \label{fig:three graphs}
\end{figure}

Recent work has investigated the potential for applying the above training method to the layout of FAST prototype telescopes located at the TA Black Rock Mesa site (FAST@TA). These three telescopes are co-located, only allowing for monocular reconstruction of events. Initial attempts at training involved $\sim300,000$ showers with core positions contained within a circle centred on (0,0) with radius 10\,km. The telescope layout and core positions are shown in Figure \ref{fig:talayout}. Preliminary results indicate that the resolution in the reconstructed parameters is too low to be used as a first guess. The lack of stereo observation and fewer input parameters compared to previous models are believed to be the cause of the worse performance. Altering the positions of the three telescopes to allow for stereo reconstruction improves results, particularly for the geometry reconstruction. To obtain a sufficiently precise first guess of all six parameters with the standard layout, methods of utilising more information from each PMT, such as directly inputting traces into a convolutional neural network or adding additional shape information to the current inputs, are being investigated. 

Alternatively, instead of estimating all six parameters, FAST@TA may use the geometry reconstruction from the TA ground array as a proxy for the true geometry, reconstructing \Xmax{} and energy only. By adding the geometrical parameters to the input of the network, we are able to obtain greater precision in the estimates of \Xmax{} and energy with the FAST@TA setup. Aside from the increased number of inputs and fewer outputs, the training method is identical. Figure \ref{fig:mlresult} shows the model predictions on 10,000 showers not used during training. With this setup, the resolution of the first-guess is $\sim75$\,g\,cm$^{-2}$ in \Xmax{} and $\sim20$\% in energy at 40\,EeV. An evaluation of the adequacy of this resolution for the top-down reconstruction is provided in the following section.

\section{Top-down reconstruction}
Once a first estimate of the shower parameters has been obtained, the top-down reconstruction optimises the result by simulating showers with parameters in the vicinity of the first guess and comparing the noise-free simulated traces to the data. Specifically, the best-fit parameters, $\vec{a}$, are those which maximize the log-likelihood function
\begin{equation}
    \ln\mathcal{L}\left(\vec{x}|\vec{a}\right)=\sum_k^{N_{\textrm{pix}}}{\sum_i^{N_{\textrm{bins}}}{P_k\left(x_i|\vec{a}\right)}}
    \label{eqn:like}
\end{equation}
where $P_k(x_i|\vec{a})$ is the probability of measuring a signal of $x_i$ photo-electrons in the $i^{\textrm{th}}$ time bin of pixel $k$. This probability is dependent on the expected value of the signal in each bin, which itself depends on the input shower parameters and background noise. In practice, $-2\ln\mathcal{L}$ is minimised using Minuit2 in ROOT\footnote{https://root.cern}, with the uncertainties on each parameter estimated from the 1$\sigma$ contours in the likelihood function. The advantages of the top-down reconstruction over relying solely on a machine learning model or a pre-simulated library of events is that time-dependent effects, such as the atmospheric conditions or telescope configuration, can be easily accounted for. Details of the simulation can be found in \cite{Thomas-Albury:2020thesis} 

\begin{figure}[t]
    \centering
    \begin{subfigure}[b]{0.49\textwidth}
         \centering
         \includegraphics[width=\textwidth]{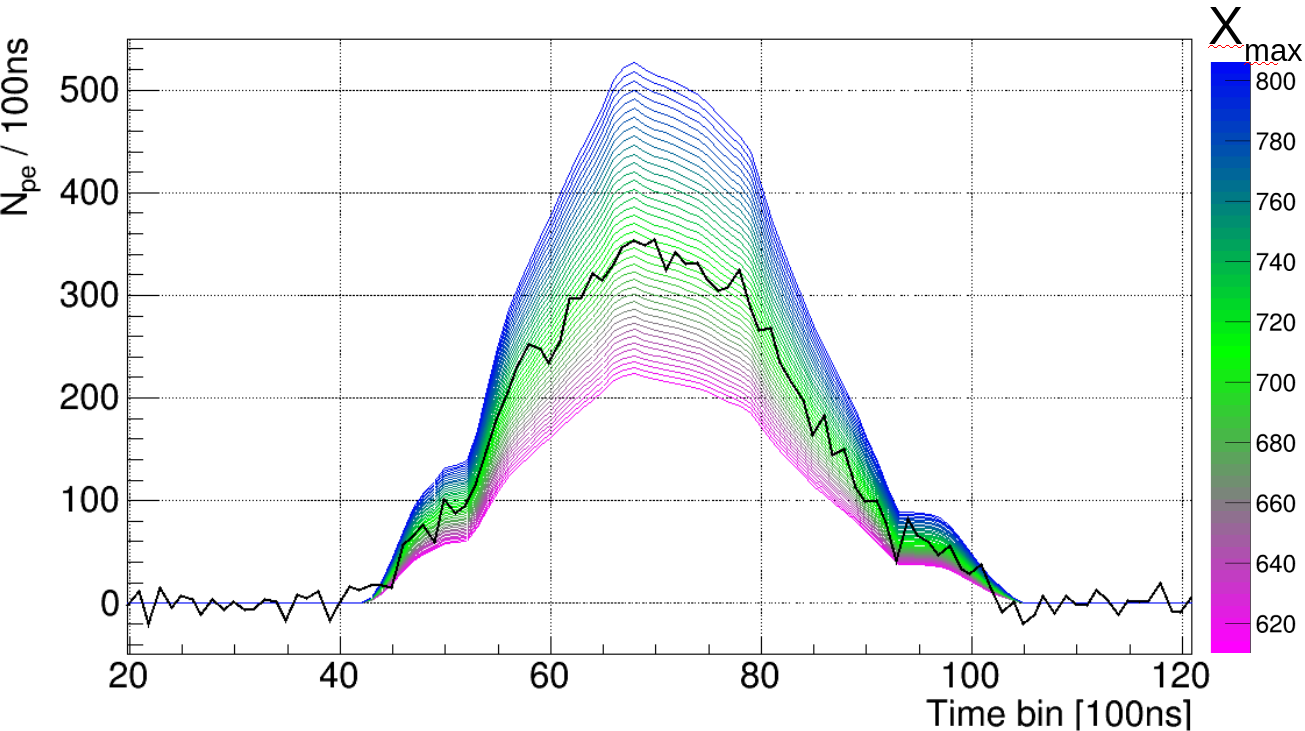}
         \caption{}
         \label{fig:tdexamp}
    \end{subfigure}
    \begin{subfigure}[b]{0.49\textwidth}
         \centering
         \includegraphics[width=\textwidth]{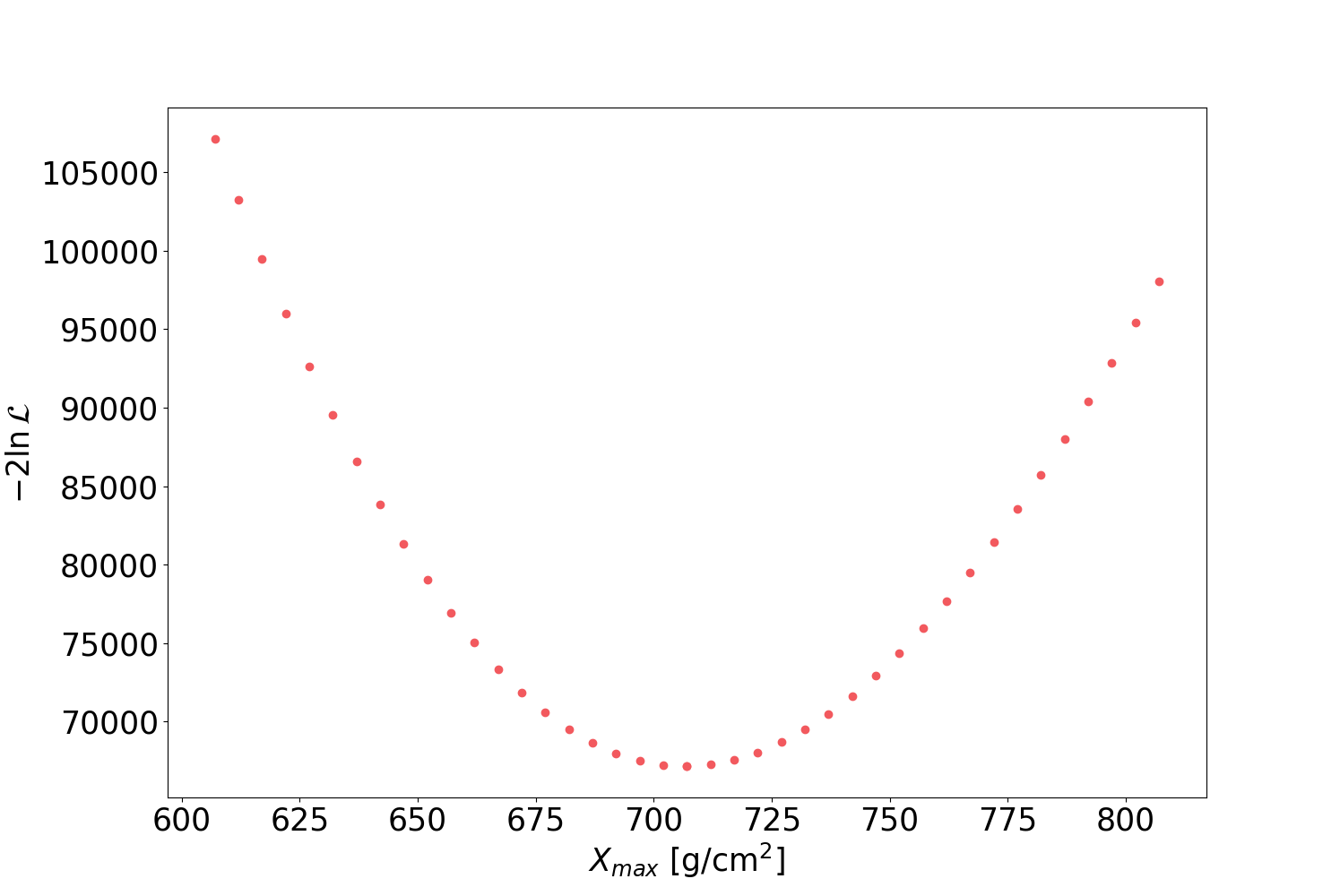}
         \caption{}
         \label{fig:like}
    \end{subfigure}
    \caption{(a) Illustration of the top-down reconstruction fitting the \Xmax{} parameter of a simulated shower with true \Xmax{} $=$ 700\,g\,cm$^{-2}$. The black line is the data, whilst each colored line represents a simulated shower with a different value of \Xmax{}. The simulated trace which ``best matches the data" (maximises the likelihood) is chosen. (b) The corresponding values of -2ln$\mathcal{L}$ for each of the simulated showers. In this case, 707\,g\,cm$^{-2}$ is the best fit value.}
    \label{fig:topdown}
\end{figure}

\begin{figure}[t]
    \centering
    \includegraphics[width=0.40\textwidth]{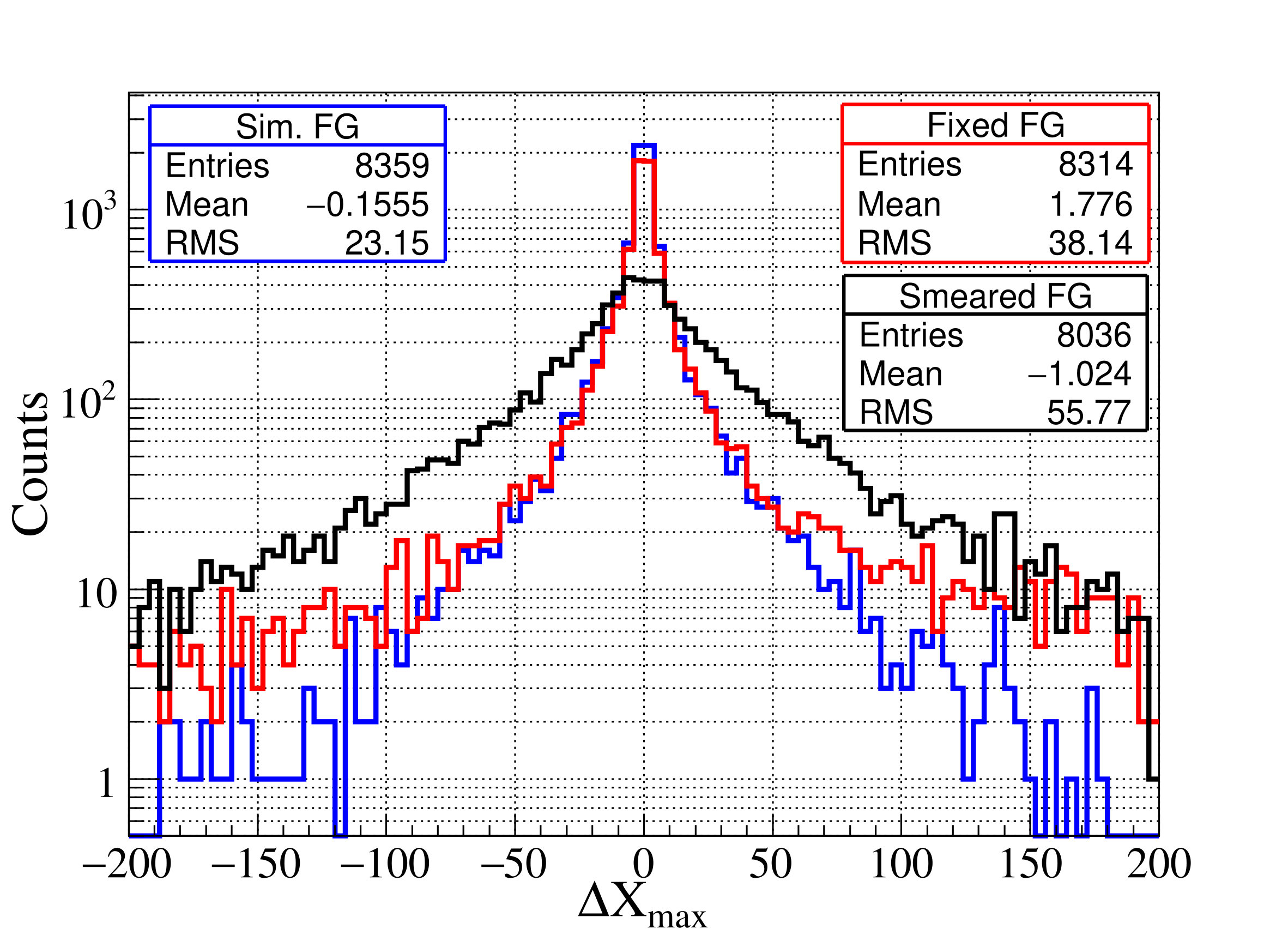}
    \includegraphics[width=0.40\textwidth]{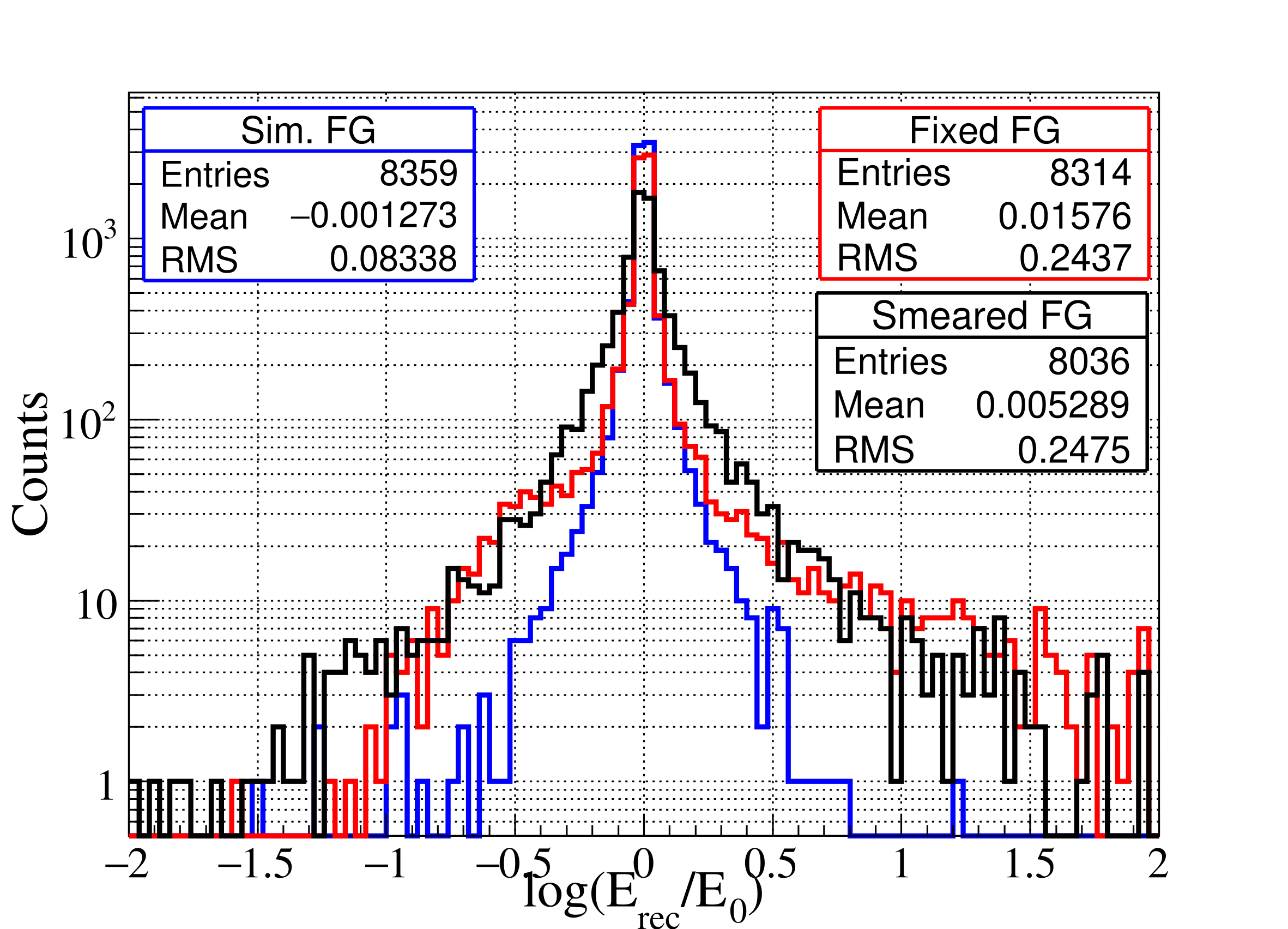}
    \caption{Comparison of the top-down reconstruction performance for FAST@TA using different initial guesses. ``FG" stands for `first-guess'.}
    \label{fig:tdFGresults}
\end{figure}

To illustrate the procedure in a simple case, Figure \ref{fig:tdexamp} shows the top-down reconstruction fitting the \Xmax{} parameter for a simulated shower (only one PMT shown). Figure \ref{fig:like} is a plot of the corresponding likelihood values.  
Similar to the machine learning step, the top-down reconstruction has been well-tested for a triangular array of FAST stations. However, with data taking progressing well at both the Auger and TA sites, it is important to understand our expected performance and be able to accurately reconstruct events with the current prototypes. To this end, we have simulated 10,000 showers with the FAST@TA arrangement and tested the top-down reconstruction of \Xmax{} and energy with varying initial guesses. Note the simulated shower parameters and core positions are as described in the machine learning section. The results are shown in Figure \ref{fig:tdFGresults}. The blue histogram uses a first guess identical to the simulated shower parameters, whilst the red histogram uses geometry identical to the simulated values but with a fixed first guess of \Xmax{} $=850$g\,cm$^{-2}$ and $E=10^{19.5}$\,eV. The black histogram uses geometry values randomly shifted from truth using the typical resolution of the TA ground array plus randomly shifted values of \Xmax{} and energy using the expected resolution of the neural network introduced in Section \ref{sec:ml}. 
\section{Reconstruction of data with FAST@TA}
\begin{figure}[h]
    \centering
    \includegraphics[width=0.35\textwidth]{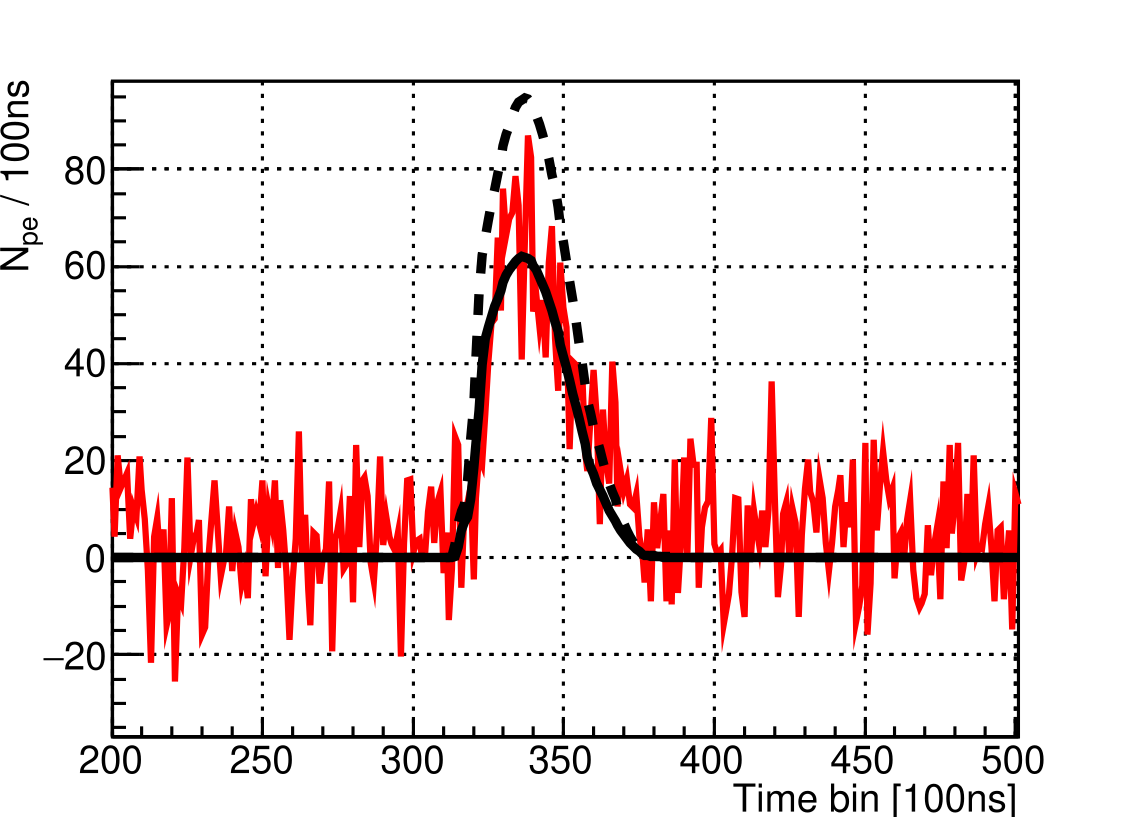}
    \includegraphics[width=0.35\textwidth]{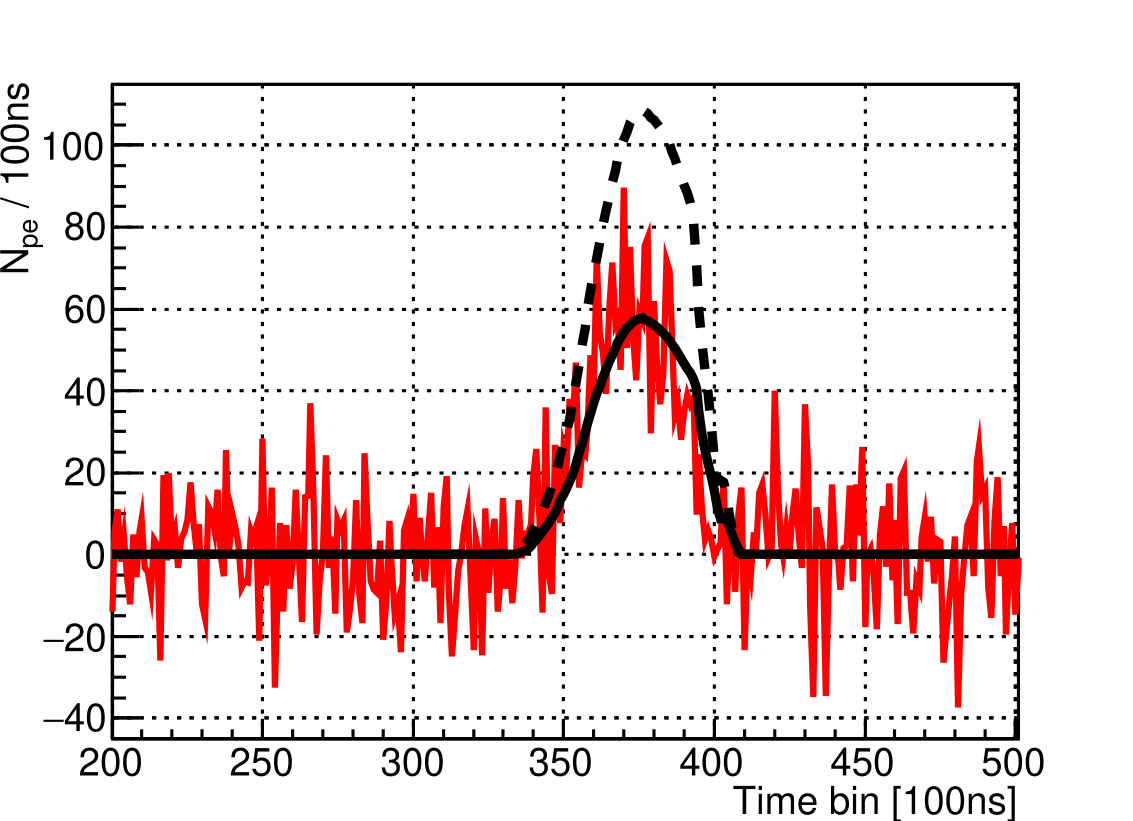}
    \caption{Top-down reconstruction fit to the significant PMT signals from event ID 2. The dashed line shows the simulated trace using the TA parameters as the first guess. The solid line shows the final result of the fitting procedure.}
    \label{fig:eventexample}
\end{figure}

In \cite{FAST:2021toshi} FAST@TA data up until September 2019 was analysed for coincidences with the TA fluorescence telescopes, with a total of 21 events found above 10$^{18}$\,eV. Here we show preliminary results of the analysis of data from October 2019 to February 2023 using the top-down reconstruction to estimate \Xmax{} and energy. In this period, we find 11 events in coincidence with the TA monocular reconstruction with TA reconstructed energies above 10$^{18}$\,eV. Using the TA reconstruction values as a first guess, 7 events were found to give reconstructions which reasonably matched the measured traces. Failed reconstructions are likely due to poor first guess values, particularly for the shower geometry. Table \ref{tab:tdresults} shows a comparison of the TA and FAST reconstructed measurements. In cases where the initial geometry didn't provide an adequate first guess, the top-down reconstruction was allowed to fit all parameters. These events are marked with a `*'. Events marked with `**' use the TA hybrid reconstruction values as a first guess. As an example, the reconstruction of event 2 is shown in Figure \ref{fig:eventexample}. 

\begin{table}[h]
    \centering
    \begin{tabular}{c|c||c|c|c|c|}
         ID & Event time & \multicolumn{2}{|c|}{TA FD Mono (Prelim.)} & \multicolumn{2}{|c|}{FAST (Prelim.)} \\
         \hline& & Energy (EeV) & \Xmax (gcm$^{-2}$) & Energy (EeV) & \Xmax (gcm$^{-2}$) \\
         \hline\hline
         1 & 2019/10/25 04:23:52 & 6.31 & 793 & 3.67$\pm$0.19 & 728$\pm$27 \\
         2* & 2020/01/28 08:20:44 & 3.02 & 865 & 1.7$\pm$0.3 & 816$\pm$49 \\
         3 & 2020/01/28 11:13:17 & 1.91 & 478 & 1.44$\pm$0.1 & 439$\pm$10 \\
         4 & 2022/11/25 09:24:16 & 1.66 & 646 & 1.54$\pm$0.13 & 384$\pm$13 \\
         5** & 2022/11/26 04:42:03 & 8.13 & 771 & 6.6$\pm$0.9 & 509$\pm$26 \\
         6 & 2023/02/17 05:13:36 & 1.55 & 561 & 1.29$\pm$0.16 & 533$\pm$19 \\
         7* & 2023/02/20 08:15:51 & 1.78 & 867 & 3.7$\pm$0.6 & 375$\pm$30 \\
    \end{tabular}
    \caption{Top-down reconstruction results for real events measured by FAST@TA.}
    \label{tab:tdresults}
\end{table}

\section{Summary}
The FAST reconstruction process is a novel method for determining the parameters of primary cosmic rays, utilising both machine learning and maximum likelihood based techniques.
We have reported on the details of these methods and determined realistic performance estimates for the reconstruction of \Xmax{} and energy with FAST@TA. We have also analysed new data from FAST@TA post 2019. Improvements to the machine learning aspect of the reconstruction, which allow for all six shower parameters to be accurately estimated with fewer telescopes, are being investigated.

\small %% Added by Toshihiro to fit 8 pages.
\acknowledgments{This work was supported by JSPS KAKENHI Grant Number 21H04470, 20H05852, 18KK0381, 18H01225, 15H05443. This work was supported by JST, the establishment of university fellowships towards the creation of science technology innovation, Grant Number JPMJFS2138. This work was partially carried out by the joint research program of the Institute for Cosmic Ray Research (ICRR) at the University of Tokyo. This work was supported in part by NSF grant PHY-1713764, PHY-1412261 and by the Kavli Institute for Cosmological Physics at the University of Chicago through grant NSF PHY-1125897 and an endowment from the Kavli Foundation and its founder Fred Kavli. The Czech authors gratefully acknowledge the support of the Ministry of Education, Youth and Sports of the Czech Republic project No. LTAUSA17078, CZ.02.1.010.0/17\_049/0008422, LM 2023032 and the support of the Czech Academy of Sciences and Japan Society for the Promotion of Science within the bilateral joint research project with Osaka Metropolitan University (Mobility Plus project JSPS 21-10). The Australian authors acknowledge the support of the Australian Research Council, through Discovery Project DP150101622. The authors thank the Pierre Auger and Telescope Array Collaborations for providing logistic support, part of the instrumentation to perform the FAST prototype measurement, and productive discussions.}

\bibliography{main}
\bibliographystyle{JHEP}
%% Full authors list (ONLY FOR COLLABORATIONS)
%\clearpage
%\section*{Full Authors List: \Coll\ Collaboration}
%
%\noindent \textbf{Note comment afterwards:} Collaborations have the possibility to provide an authors list in xml format which will be used while generating the DOI entries making the full authors list searchable in databases like Inspire HEP. \\
%
%\scriptsize
%\noindent
%first.author$^1$, 
%second.author$^2$, 
%third.author$^3$ % .... more names
%and 
%last.author$^{n}$ \\
%
%\noindent
%$^1$first.affiliation.
%$^2$second.affiliation. % .... more affiliation
%$^{m}$last.affiliation.

\end{document}